\documentclass{aa}

\usepackage{graphicx}
\usepackage{txfonts}
\usepackage{scrextend}
\usepackage{hyperref}
\begin{document} 
\newcommand*{\rebound}{\textsc{REBOUND}\xspace}
\newcommand*{\horizons}{\textsc{horizons}\xspace}
\defcitealias{H+16}{H+16}

    \title{The imprint of the MOND external field effect\\ on the Oort cloud comets aphelia distribution}

   \author{R. Pau\v{c}o, J. Kla\v{c}ka
          }

   \institute{Faculty of Mathematics, Physics and Informatics, Comenius University in Bratislava,
    Mlynsk\'{a} dolina, 842 48 Bratislava\\
              \email{pauco@fmph.uniba.sk}
             }


 
\abstract
{Milgromian dynamics (MD or MOND) is a promising physical description excelling especially in galaxies. When formulated as a modified gravity theory, it leads to the so called external field effect (EFE). In the case of the solar system this means that bodies orbiting the Sun are influenced, beyond its tidal effect, by the external gravitational field of the Galaxy with magnitude $\sim2\times10^{-10}$ m s$^{-2}$ and time-varying direction. Aphelia of intermediate outer Oort cloud (OC) comets ($30<X<60,~X\equiv 10^{6}/a[\text{au}]$, where $a$ is semimajor axis) are distributed non-uniformly on the celestial sphere, showing an apparent concentration around a great circle centered at Galactic longitudes $L=-45$ and 135 deg. Such non-uniformity is beyond that attributable to the classical injectors of comets, stellar encounters and the Galactic tides, as well as the expected observational biases.}
{We investigated a hypothesis that the great circle concentration of aphelia is a consequence of the long-term action of EFE in the framework of MD.} 
{We considered exclusively quasi-linear MOND (QUMOND) theory. We built our model of the OC in MD on an analytical approximation of the QUMOND potential for a point mass in the dominant external field of constant magnitude. The model is well applicable at heliocentric distances $r\gtrsim10~000$ au.  Constraints on the strength of EFE found by the analysis of the Cassini radio-tracking data were taken into account.}
{We demonstrated characteristic imprint of the EFE on the distribution of aphelia of candidate outer OC comets that migrated down to $r=10~000$ au. By both analytical and numerical calculations, we showed that the combined effect of EFE and the Galactic tides could qualitatively account for the characteristic features seen in the observed distribution of aphelia of the outer OC comets.}
{}


   \maketitle
%

\section{Introduction}\label{sec:intro}

Spatial orientation of original orbits (i.e. before entering the planetary zone) of new Oort cloud (OC) comets can teach us about the dynamical processes that delivered them from the OC into the inner solar system.
The standard picture is that the Galactic tides and impulses from the stars passing by the OC cooperate in decreasing of perihelion distances of OC bodies, making them eventually observable as new comets \citep{HT86,R+08}. The role of passing stars is important especially in the repopulating of the tidal infeed trajectories \citep{R+08}. Aphelia of new outer Oort cloud (OOC) comets (having semimajor axes, $a$, $a>$ 10 000 au) are distributed non-uniformly on the celestial sphere.
In the Galactic coordinates, the latitudes of aphelia, $B$, are depleted in the equatorial and the polar regions, a feature first noted by \citep{Del87}.
This is in accord with what is expected when the Galactic tides make the comets discernible \citep{Tor86}.
When we restrict ourselves to comets with accurately known original orbital energies (class 1A orbits of \citealp{MW08}), $\vert\sin B\vert$ distribution\footnote{For random $B$s, the $\vert\sin B\vert$ distribution would be flat.} peaks around 0.5 \citep{MW11}. The position of the peak agrees well with the one given by the combined model (stellar encounters + Galactic tides) of \cite{R+08}; see their Fig. 5.

\cite{M+99} identified an excess of intermediate $(30<X<60,~X\equiv 10^{6}/a[\text{au}])$ OOC comets aphelia lying on the celestial sphere close to a great circle centered at Galactic longitudes, $L$, $L=-45$ and 135 deg. The great circle excess is still present also in newer data \citep{MW11}. The excess is the most apparent for high-quality orbits, i.e. for the  comets about which we are most certain that they are the first-time entrants into the planetary region.
The non-uniformity in $L$ seems to be beyond that attributable to the known observational biases \citep{HE02} and the selection induced by the conventional cometary injection mechanisms \citep{HT86,R+08,M+99}. 
A potential explanation put forward by \cite{M+99} is that the excess is induced by a perturbation coming from an unseen massive body 
with mass of few Jupiter masses, interacting with comets impulsively at a mean distance of $\approx 25~000$ au. There is not much room for such perturber in the solar system as implied by the analysis of the Wide-field Infrared Survey Explorer (WISE) data \citep{Luh14}.

\cite{FBJ14} stressed that the aphelia cluster around the preferred plane the normal to which is the solar apex. The authors argue that the non-uniform $L$ distribution is an effect of the solar apex motion and that a few strong stellar encounters with proper geometry could be able to produce it. Although, they were not be able to identify such past encounter in their sample of past stellar encounters experienced by the Sun; see also \cite{FBJ15}.


Modified Newtonian dynamics (MOND) (\citealp{Mil83}; \citealp{FM12} for a review), is a very successful paradigm for galaxies. It predicted existence of the radial acceleration relation \citep{McG+16}, which generalizes well-known systematic properties of galaxies -- the mass discrepancy-acceleration relation \citep{San90,McG04} and the baryonic Tully-Fisher relation \citep{McG+00}. In galaxies everything happens as if the gravitation was to be calculated according to Milgrom, not Newton with aid of dark matter \citep{Kro+10,K12,FM12,McG+16}. More than 30 years ago, Milgrom proposed that dynamics in the low acceleration regime -- when accelerations are much smaller than some fundamental acceleration $a_{0}$, $a_{0}\sim10^{-10}$ m s$^{-2}$ -- departs from Newtonian dynamics. A particle in the low acceleration regime subject to influence of a spherically symmetric and isolated mass distribution, moves with acceleration of magnitude 
\begin{eqnarray}\label{MOND0}
g=\sqrt{a_{0}g_{N}}~,
\end{eqnarray}
where $g_{N}$ is the expected Newtonian gravitational acceleration due to that mass distribution. Eq. (\ref{MOND}) is the cornerstone of MOND, meaning that MOND is space-time scale invariant in the low acceleration regime \citep{Mil09}, contrary to Newtonian dynamics or general relativity. To ensure better clarity, we call fully-fledged classical theories incorporating MOND -- Milgromian dynamics (MD).

As first noted by \cite{Mil86}, the gravitational potential seen by objects in the OC in MD is not spherically symmetric but dilated along the direction of the external field of the Galaxy. The dilation of the potential is beyond that introduced by the tidal effects. Milgrom speculated that this, plus the enhanced gravitational field of the Sun in MD, could have some impact on our concept of the Oort cloud.
Recently, impact of MD on the Oort cloud \citep{PK16} and the Kuiper belt \citep{Pau17} was studied. When constraints on the theory based on the Cassini radio-tracking data \citep{H+14,H+16} (the latter hereafter referred to as \citetalias{H+16}) are applied, the Oort cloud has a similar size and the planetary barrier operates similarly to its Newtonian counterpart \citep{PK16}. But the delivery of comets could be substantially influenced by the so called external field effect (EFE). EFE could also produce enough torque to detach the orbits of Sedna and 2012 VP$_{113}$ \citep{Pau17}. 
EFE brings in a preferred direction, the time-varying direction of the external field of the Galaxy, which in MD does not decouple from the internal dynamics \citep{Mil83,FM12}. Hence, there are preferred regions in space where EFE causes larger angular momentum changes of OC objects. This in turn could lead to the observed non-uniformity in the $(L,B)$ distribution of the OOC comets. An imprint of the EFE would be the most easily observed in the regime where tides do not dominate over EFE, i.e. on orbits of intermediate and low-$a$ comets. In this paper, we investigate a hypothesis that the great circle concentration (GCC) of aphelia identified by \cite{M+99} is caused by the EFE of MD. 

\section{Distribution of outer Oort cloud comets aphelia}\label{sec:data4}

\begin{figure}
\begin{center}
\resizebox{\hsize}{!}{\includegraphics{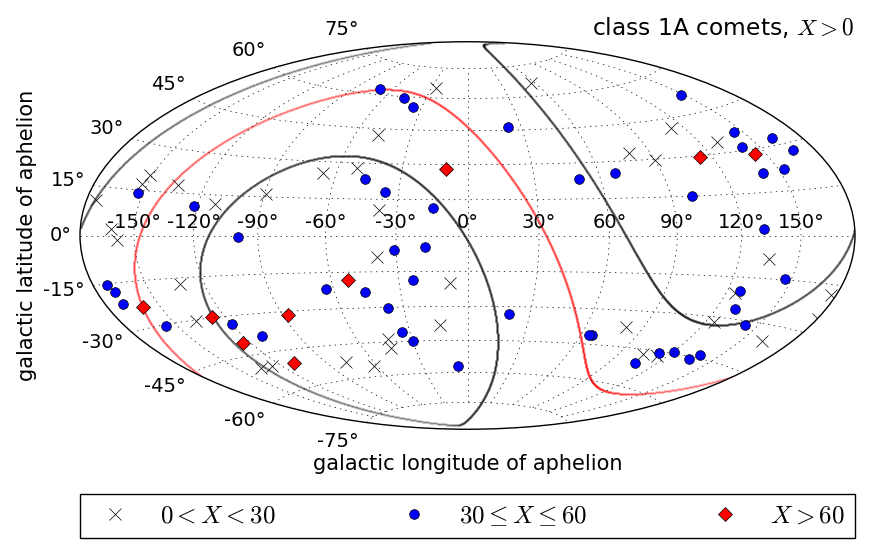}}
\caption{Aphelion distribution of the highest-quality (1A) orbits of the bounded OOC comets. The data were retrieved from the 17th Catalogue of Cometary Orbits \citep{MW08}. This is an equal-area Hammer-Aitoff projection in the Galactic coordinates. 
The comets are discerned according to their energies and divided into three groups: i) those with $0<X<30$ (crosses), ii) $30\leq X\leq 60$ (circles), iii) $X>60$ (diamonds).
The comets in the region with longitudes between -180 and -120 deg and latitudes between 0 and 30 deg are probably a residual part of a comet shower. The equatorial plane (solid red line) and declination levels of $\pm30$ deg (solid black lines) are indicated. The north-south bias in the equatorial coordinates of aphelia is not so eminent in the high-quality data.}
\label{img:data}
\end{center}
\end{figure}

The aphelia distribution for the highest-quality (1A) orbits of OOC comets is shown in Fig. \ref{img:data}. This is an equal-area Hammer-Aitoff projection in the Galactic coordinates. The data were taken from \cite{MW11}, and originally from the 17th Catalogue of Cometary Orbits \citep{MW08}. The comets are discerned according to their energies and divided into three groups: i) those with $0<X<30$ ($a~\gtrdot~33~300$ au), ii) $30\leq X\leq 60$, iii) $X>60$ ($a~\lessdot~16~700$ au). We consider only bounded $(X>0)$ comets.
Comets with aphelion longitudes, $L$, between -180 and -120 deg and latitudes, $B$, between 0 and 30 deg, are probably a residual part of a weak comet shower \citep{B+83,M+98}. This shower is evidenced mainly by the low-$X$ (high-$a$) comets.

A strong north-south (NS) bias in the equatorial coordinates of aphelia is prevalent in the cometary catalogue as a whole \citep{HE02}. The NS bias is a consequence of the simple fact that more observations have been carried out from the northern than from the southern hemisphere in the past. However, when we restrict ourselves to the 1A category orbits, the NS bias is not so apparent. This can be seen in Fig. \ref{img:data}. There are 15 1A comets with aphelion declination higher than 30 deg north, while 19 1A comets have declination higher than 30 deg south. We note that the comets discovered in the north have generally perihelia in the north and vice versa \citep{HE02}; perihelion in the north means aphelion in the south.
Interestingly, there appear to be a preference for the southern aphelia not in the equatorial but in the Galactic coordinates when we look at $\vert B\vert>30$ deg region. If it was not for this NS asymmetry in the Galactic coordinates, the data would look more or less uniformly distributed for $\vert B\vert>30$ deg, with polar holes for $\vert B\vert>70$ deg.

The cometary data are burdened with various selection effects. These include seasonal and diurnal biases, directional and orbital biases, meaning that comets in some directions on the sky or in some specific orbits are more likely to be discovered, as well as some sociological biases \citep{HE02}. Many of these biases cancel each other, while none of them is able to induce systematics of the observed aphelion distribution \citep{HE02}. 

\begin{figure}
\begin{center}
\resizebox{\hsize}{!}{\includegraphics{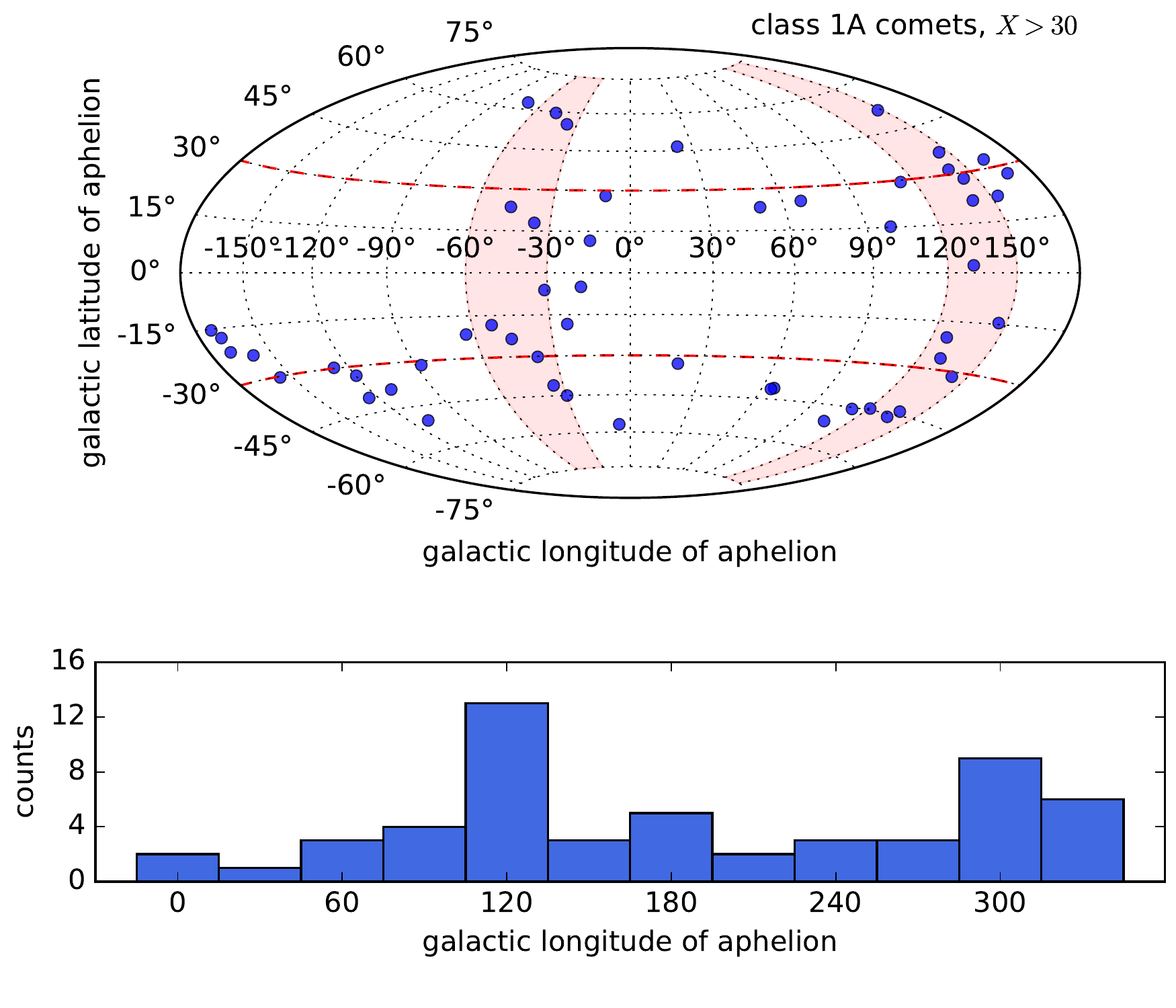}}
\caption{\textbf{Top.} Aphelion distribution of the highest-quality (1A) orbits of the OOC comets having $X>30$. The data were retrieved from the 17th Catalogue of Cometary Orbits \citep{MW08}. This is an equal-area Hammer-Aitoff projection in the Galactic coordinates. Three comets having $X>30$ in the region with longitudes between -180 and -90 deg and latitudes between 0 and 30 deg were omitted, as these could be associated with a comet shower. The overpopulated great circle identified by \cite{M+99} is highlighted in red. The equatorial zone, where $\vert B\vert<30$ deg, and where the tides have lesser effect and the non-uniformity is the most apparent, lies between the dashed lines. \textbf{Bottom.} Histogram distribution of the cometary Galactic longitudes of aphelia. Two prominent peaks emerge around longitudes 135 and 315 (-45) deg.}
\label{img:data_tocomp}
\end{center}
\end{figure}

We show aphelion positions in the Galactic coordinates and the histogram distribution of $L$ for $X>30$ 1A-class comets only in Fig. \ref{img:data_tocomp}. The two comets in the region with longitudes between -180 and -120 deg and latitudes between 0 and 30 deg were discarded as these are probably associated with a comet shower \citep{B+83,M+98}. We discarded also one additional intermediate OOC lying close, in terms of $(L,B)$, to the previously discarded couple, and whose aphelion is nearly in the plane of the Galaxy. Therefore, this comet is a good candidate for a comet made discernible due to a stellar encounter. The overpopulated great circle of aphelia identified by \cite{M+99} is apparent in the figure, especially in the Galaxy's equatorial region. 
These are the puzzling data we aim to explain with the aid of MD.


\section{Model and analytic calculations}\label{sec:model4}

Two classical modified gravity theories, giving arise to MOND phenomenology in galaxies, are in circulation -- fully non-linear aquadratic-lagrangian theory (AQUAL; \citealp{BM84}) and quasi-linear MOND (QUMOND; \citealp{Mil10}).
To ensure better clarity, we call such fully-fledged classical theories -- Milgromian dynamics (MD).
AQUAL and QUMOND coincide for spherically symmetric problems \citep{BM84,Mil10}, giving the equation of motion in the form of the simple MOND formula \citep{Mil83}:
\begin{eqnarray}\label{MOND}
{\bf g} = \nu( g_{N}/a_{0}){\bf g}_{N}~,
\end{eqnarray}
where ${\bf g}_{N}=-\nabla\Phi_{N}$ is the expected Newtonian gravitational acceleration, $g_{N}$ is its norm, $\nu(\eta_{N})$, $\eta_{N}\equiv g_{N}/a_{0}$, is a modification factor, the so called interpolating function, fulfilling behaviour $\nu(\eta_{N})\rightarrow 1$ for $\eta_{N}\gg1$ (Newtonian limit) and $\nu(\eta_{N})\rightarrow \eta_{N}^{-1/2}$ for $\eta_{N}\ll1$ (deep-MD limit -- Eq. (\ref{MOND0})), and $a_{0}\sim 10^{-10}$ m s$^{-2}$ is a new constant of nature.
The two theories are distinct outside of the spherical symmetry, although they often give a similar qualitative picture.

We build our model of the OC on the simpler QUMOND theory.
In QUMOND, the governing gravitational potential, $\Phi$, is given by \citep{Mil10}:
\begin{eqnarray}\label{QUMOND4}
\nabla^{2}\Phi~=~\nabla\cdot\left[\nu\left(\frac{\vert\nabla\Phi_{N}\vert}{a_{0}}\right)\nabla\Phi_{N}\right]~.
\end{eqnarray}
The problem can be reformulated with aid of phantom matter density (algebraically calculable from the Newtonian potential $\Phi_{N}$) \citep{Mil10}, hence one does not have to solve Eq. (\ref{QUMOND4}) but two linear Poisson equations instead (see e.g. Sect. 6.1.3 of \cite{FM12} for details).

In our exploratory analysis of the problem in MD, we will restrict ourselves to the region of the Oort cloud where the external field from the Galaxy dominates over the internal one from the Sun. 
This means heliocentric distances larger than $\sqrt{GM/g_{N,ext}}$, where $G$ is gravitational constant, $M$ is mass of the Sun, and $g_{N,ext}$ is magnitude of the external Newtonian gravitational acceleration.
Under such condition, one can avoid the direct numerical solution of the Poisson equation. The governing QUMOND potential can be in this case well approximated by \citep{BZ16}:
\begin{eqnarray}\label{QUMOND_EFE}
\Phi &=& -\frac{GM\nu_{ext}}{r}\left(1+\frac{K_{0}}{2}\sin^{2}\theta\right) ~,
\end{eqnarray}
where $\nu_{ext}\equiv\nu(\eta_{N,ext})$, $\eta_{N,ext}\equiv g_{N,ext}/a_{0}$,
\begin{eqnarray}\label{dlnmu}
K_{0}\equiv\frac{d\ln\nu}{d\ln\eta_{N}}\bigg\vert_{\eta_{N}=\eta_{N,ext}}~,
\end{eqnarray}
and $\theta=\arccos\left(\widehat{{\bf g}}_{N,ext}\cdot \widehat{{\bf r}}\right)$, $\widehat{{\bf g}}_{N,ext}={\bf g_{N,ext}}/g_{N,ext}$, $\widehat{{\bf r}}={\bf r}/r$, i.e. $\theta$ is the angle between the direction of the external Newtonian field and the direction towards the Sun. 
$\eta_{N,ext}$ is to be calculated from $\eta_{N,ext}\nu(\eta_{N,ext})=\eta_{ext}$, where $\eta_{ext}$ can be estimated for instance from the solar motion. We use a simple model of the Sun's motion -- the Sun orbiting in a circular orbit, lying in the Galactic midplane, with angular velocity $\Omega_{0}=V_{0}/R_{0}$. Then, $\eta_{ext}=V^{2}_{0}/(a_{0}R_{0})$ is the centripetal acceleration of the Sun in units of $a_{0}$.

$\Phi=\Phi(t)$, as the direction of the external field, coinciding with the direction towards the Galactic center, varies with time. We use
\begin{eqnarray}\label{variest}
e_{x} & = & \cos(\Omega_{0}t)~, \nonumber \\
e_{y} & = & \sin(\Omega_{0}t)~, \nonumber \\
e_{z} & = & 0~, \nonumber \\
\Omega_{0} & < & 0~.
\end{eqnarray}
The Sun orbits the Galactic center clockwise from the north-Galactic-pole perspective.

For an OC body orbiting in the Galactic midplane $(B=0)$, Eqs. (\ref{QUMOND_EFE}), (\ref{variest}), and the Gauss planetary equations lead to secular variations:
\begin{eqnarray}\label{dddt}
\left\langle \frac{da}{dt} \right\rangle & = & 0~, \nonumber \\
\left\langle \frac{de}{dt} \right\rangle & = & \mu^{1/2}K_{0}\nu_{ext}\frac{\sin [2(\vert\Omega_{0}\vert~t+L)]}{2a^{3/2}}F(e)~ \nonumber \\
\left\langle\frac{dL}{dt}\right\rangle & = & - \mu^{1/2}K_{0}\nu_{ext}\frac{\cos[2(\vert\Omega_{0}\vert~t+L)]}{2a^{3/2}}G(e)~,
\end{eqnarray}
where $a$ and $e$ are semimajor axis and eccentricity of the body's orbit, and $F(e)$ and $G(e)$ are functions involving only eccentricity; details can be found in appendix \ref{sec:app4}. We have averaged over true anomaly. Eqs. (\ref{dddt}) form a closed set of equations for $a$, $e$, and $L$.
As $\left\langle da/dt \right\rangle=0$, the secular change $\left\langle dq/dt \right\rangle$, where $q$ is perihelion distance of the orbit, reads 
\begin{eqnarray}\label{dqdt}
\left\langle \frac{dq}{dt} \right\rangle=-a\left\langle \frac{de}{dt} \right\rangle~. 
\end{eqnarray}
As $K_{0}\nu_{ext}<0$ and $F(e)<0$ for $e\in\langle 0,1)$, $\left\langle dq/dt \right\rangle$ is negative and extremal when $\vert\Omega_{0}\vert t + L=\pi/4$ or $5\pi/4$. Moreover, in such case $\left\langle dL/dt\right\rangle=0$. If we have $\vert\Omega_{0}\vert t=(2C-1)\pi/2$, where $C$ is an integer, $q$ decreases the most rapidly on a great circle crossing $L=-45$ and 135 deg, same one as the one indicated by the observations of the intermediate OOC comets. However, this does not correspond to the present-day orientation of the external field ($\vert\Omega_{0}\vert t=2C\pi$ does). 
In any case, as of the non-stationarity of EFE and combination of the action of EFE and tides, we opt to do some numerical experiments; see Sect. \ref{sec:sim}. 

Eqs. (\ref{dddt}) and (\ref{dqdt}) lead to $\left\langle dq/dt \right\rangle\propto a^{-1/2}$, i.e. $\left\langle dq/dt \right\rangle$ is a decreasing function of $a$.
The fact that we observe preference for some values of $L$ in the intermediate comets sample, while no preferred values for comets with higher $a$, can be explained by the dominance of the Galactic tide as a comet injector at larger $a$. For angular momentum magnitude $J$ changed due to the effect of the Galactic tide, we have $\left\langle dJ/dt \right\rangle\propto a^{7/2}$ (e.g. \citealp{Tor86}), and we have averaged over true anomaly again. $J=\sqrt{2GMq}$ holds approximately for near-parabolic comets, hence we come to $\left\langle dq/dt \right\rangle\propto a^{4}$. 


\subsection{Tides}\label{sec:tides}

Provided that the Galactic mass is distributed axisymmetrically and the Galactic rotation curve is flat at the Sun's location, the Newtonian tidal acceleration of an OC object orbiting the Sun is \citep{HT86,Fou04}:
\begin{eqnarray}\label{tides}
\ddot{x}_{tides} & = & \Omega^{2}_{0}x'\cos(\Omega_{0}t)+\Omega^{2}_{0}y'\sin(\Omega_{0}t)~, \nonumber \\
\ddot{y}_{tides} & = & \Omega^{2}_{0}x'\sin(\Omega_{0}t)-\Omega^{2}_{0}y'\cos(\Omega_{0}t)~, \nonumber \\
\ddot{z}_{tides} & = & -4\pi G \varrho_{0} z~,
\end{eqnarray}
where $x'=x\cos(\Omega_{0}t)+y\sin(\Omega_{0}t)$, $y'=-x\sin(\Omega_{0}t)+y\cos(\Omega_{0}t)$, $\Omega_{0} < 0$, and $\varrho_{0}$ is local mass density in the solar neighborhood. 
To keep the model in MD simple, we add the tidal acceleration of Eq. (\ref{tides}) to the acceleration calculated from Eq. (\ref{QUMOND_EFE}). The hint that this might be a good approximation comes from the fact that in QUMOND the problem can be formulated in the Newtonian manner with the aid of phantom mass calculated from the distribution of stars and gas in the Galaxy \citep{Mil10,FM12}. Moreover, at the Sun's location, an enhancement of the gravity due to MD is low, or equivalently, there is not an obvious need for dark matter at the Sun's location.

\subsection{Interpolating function and Galactic parameters}\label{sec:IF}

The Cassini radio-tracking data put constraints on the MD interpolating function families as these are related to the strength of EFE which is the dominant effect of MD in the planetary region of the solar system \citepalias{H+16}. Interpolating functions come in pairs with values of $a_{0}$, based on the rotation curve fits of galaxies \citepalias{H+16}.
As a representative of the allowed interpolating functions we choose
\begin{eqnarray}\label{IF}
\overline{\nu}_{\alpha}(\eta_{N}) & = & \left[ 1-\exp(-\eta_{N}^{\alpha})\right]^{-1/(2\alpha)} + \left(1-\frac{1}{2\alpha}\right)\exp(-\eta_{N}^{\alpha})~, \nonumber \\
\alpha & \geq & 2
\end{eqnarray}
family.

In Table \ref{tab:IF_tab}, we show parameters $\nu_{ext}$ and $K_{0}$ of Eq. (\ref{QUMOND_EFE}) for a given set -- interpolating function, $a_{0}$, $\eta_{ext}$. Two values of the external field strength are adopted, $g_{ext}=1.9\times10^{-10}$ m s$^{-2}$ and $g_{ext}=2.4\times10^{-10}$ m s$^{-2}$, the same as in \citetalias{H+16}. In addition to $\overline{\nu}_{\alpha}$ family, defined in Eq. (\ref{IF}), with $\alpha=2$ and 3, we also list $\nu_{\alpha}$ and $\widehat{\nu}_{\alpha}$, defined as in \citetalias{H+16}, with their boundary $\alpha$s (larger $\alpha$s are allowed by the Cassini data). With $g_{ext}=1.9\times10^{-10}$ m s$^{-2}$, the region where the external field dominates over the internal one is $r\gtrsim 5500$ au, while with $g_{ext}=2.4\times10^{-10}$ m s$^{-2}$, we get $r\gtrsim 5000$ au, assuming $\overline{\nu}_{2}$ interpolating function which pairs with $a_{0}=0.815\times10^{-10}$ m s$^{-2}$ \citepalias{H+16}. We note that $\eta_{ext}-\nu(\eta_{N,ext})\eta_{N,ext}=0$, based on Eq. (\ref{MOND}), gives $\eta_{ext}\approx \eta_{N,ext}$ for all allowed interpolating functions.

The values of the angular speed of the Sun $\Omega_{0}$ used in our models with $\overline{\nu}_{2}$, which is present in Eqs. (\ref{variest}) and (\ref{tides}), are listed in Table \ref{tab:gal_tab} for a given external field strength. We determined $\Omega_{0}$  in two different ways: (i) assuming fixed $R_{0}=8.3$ kpc, we calculated $\Omega_{0}$ from $\Omega_{0}=\sqrt{g_{ext}/R_{0}}$, and (ii) by making use of the result of \cite{McMB10}, i.e. $\Omega_{0}$ the most probably lies in the interval $(3.058,3.232)\times10^{-8}$ yr$^{-1}$. In (ii), we associate the lowest $\Omega_{0}$ from this interval with $g_{ext}=1.9\times10^{-10}$ m s$^{-2}$ and the highest $\Omega_{0}$ from this interval with $g_{ext}=2.4\times10^{-10}$ m s$^{-2}$.
The table also lists the corresponding values of the circular speed of the Sun $V_{0}$. For the second free parameter in Eq. (\ref{tides}), $\varrho_{0}$, we assume $\varrho_{0}=0.1$ $M_{\odot}~\text{pc}^{-3}$ \citep{HF04} throughout the paper.

\begin{table}[]
\centering
\caption{Parameters $\nu_{ext}$ and $K_{0}$ of Eq. (\ref{QUMOND_EFE}) for different interpolating functions allowed by the Cassini data (Table 2 in \citetalias{H+16}). The leftmost column lists different interpolating functions as defined in \citetalias{H+16}. The boldfaced line is the only combination of parameters leading to a great circle concentration in $L$ similar to the observed one; see Sect. \ref{sec:results4}.
$[a_{0}]=[g_{ext}]=10^{-10}$ m s$^{-2}$.}
\label{tab:IF_tab}
\begin{tabular}{|c|c|c|c|c|c|}
\hline
IF & $a_{0}$ & $g_{ext}$ & $\eta_{ext}$ & $\nu_{ext}-1$ & $K_{0}$ \\
$[~-~]$ & & & $[~-~]$ & $[~10^{-3}~]$ & $[~10^{-2}~]$ \\
\hline
\hline
${\bf \overline{\nu}_{2}}$ & \textbf{0.815}  &  \textbf{1.9} & \textbf{2.3} & \textbf{5.04575} & \textbf{-5.31581}\\
$\overline{\nu}_{2}$ & 0.815  &  2.4 & 2.9 & 0.22264 & -0.37441 \\
\hline
\hline
$\overline{\nu}_{3}$ & 0.743  &  1.9 & 2.6 & 0.00002 & -0.00012\\
$\overline{\nu}_{3}$ & 0.743  &  2.4 & 3.2 & 0.00000 & -0.00000 \\
\hline
\hline
$\nu_{7}$  & 1.450  &  2.4 & 1.7 & 3.36631 & -2.27218\\
\hline
\hline
$\widehat{\nu}_{6}$ & 1.460  &  2.4 & 1.6 & 2.80043 & -3.46535\\
\hline
\end{tabular}
\end{table}

\begin{table}[]
\centering
\caption{Galactic parameters. We adopt two values of $g_{ext}$, the same as \citetalias{H+16} did. The first two lines represent the Galactic parameters calculated from these two values of $g_{ext}$ and fixed $R_{0}=8.3$ kpc. The last two lines make use of the result of \cite{McMB10}: $\Omega_{0}$ the most probably lies in the interval $(3.058,3.232)\times10^{-8}$ yr$^{-1}$. We associate the lowest $\Omega_{0}$ from this interval with $g_{ext}=1.9\times10^{-10}$ m s$^{-2}$ and the highest $\Omega_{0}$ from this interval with $g_{ext}=2.4\times10^{-10}$ m s$^{-2}$.
Unique parameters of our models are indicated with the dagger $(\dag)$.
We assume $\overline{\nu}_{2}$ $(\dag)$ interpolating function and $a_{0}=0.815\times 10^{-10}$ m s$^{-2}$ $(\dag)$ \citepalias{H+16} in this table. Local mass density $\varrho_{0}=0.1$ $M_{\odot}~\text{pc}^{-3}$ $(\dag)$ \citep{HF04} is assumed throughout the paper.}
\label{tab:gal_tab}
\begin{tabular}{|c|c|c|c|c|}
\hline
$g_{ext}$ & $\eta_{ext}$ $^{\dag}$ & $R_{0}$ & $V_{0}$ & $\Omega_{0}$ $^{\dag}$\\
$[10^{-10}$ m s$^{-2}]$ & $[~-~]$ & $[\text{kpc}]$ & $[\text{km~s}^{-1}]$ & $[10^{-8}~\text{yr}^{-1}]$ \\
\hline
\hline
1.9  &  2.3  &  8.3  &  221  &  2.723 \\
2.4  &  2.9  &  8.3  &  248  &  3.056 \\
\hline
1.9  &  2.3  &  6.6  &  197  &  3.058 \\
2.4  &  2.9  &  7.4  &  234  &  3.232 \\
\hline
\end{tabular}
\end{table}

\section{Simulation}\label{sec:sim}

We are interested in the past orbital evolution of OOC comets with intermediate two-body binding energies, $30 < X < 60$ ($17000\lessdot~a~\lessdot33000$ au). The non-uniformity in the $L$ distribution is the most apparent in this $X$-range. This is likely because the tides in this $X$-range are not strong enough to wash away the anisotropy induced by an unknown dynamical mechanism. Also, in the case of MD, EFE is important in this region. The OOC with $X>60$ seem more or less consistent with the trend given by the intermediate comets, though it is only 9 objects; see Fig. \ref{img:data}.

The great circle excess was previously attributed to the impulsive action of a hypothetical massive jovian planet orbiting in the OOC \citep{M+99,MW11}.
We hypothesise that the anisotropy, best seen for $\vert B\vert<30$ deg data, comes from the EFE of MD. As a benchmark Milgromian model we use the model with $\overline{\nu}_{2}$ interpolating function, $\eta_{ext}=2.3$ ($g_{ext}=1.9\times10^{-10}$ m s$^{-2}$), $\Omega_{0}=2.723\times10^{-8}$ yr$^{-1}$, and $\varrho_{0}=0.1$ $M_{\odot}$ pc$^{-3}$, see Tables \ref{tab:IF_tab} and \ref{tab:gal_tab}.

We performed several test-particle simulations in Newtonian and MD, concentrating on the particles' migration. In Newtonian simulations, gravity of the Sun and tides from the Galaxy (both radial and vertical component) as in Eq. (\ref{tides}) are taken into account. 
In Milgromian simulations, acceleration calculated from the potential in Eq. (\ref{QUMOND_EFE}) and tides from the Galaxy as in Eq. (\ref{tides}) were simply added together.
180 000 test particles were started with aphelia randomly distributed on the celestial sphere. We started particles with $X$ randomly drawn from a uniform distribution between 30 and 60 ($17000\lessdot~a\lessdot33000$ au). Perihelion distances were assigned randomly from a uniform distribution between $10^{6}$/60 and $a$ au, where $a$ is the semimajor axis of a given particle. $\omega$, $\Omega$, $\cos i$, and $M$ (we use the standard notation for the orbital elements, and the angles may be defined with respect to an arbitrary reference frame) were drawn from a flat distribution.

As applicability of Eq. (\ref{QUMOND_EFE}) is limited to the region where the gravity of the Sun is smaller than the external field $g_{ext}$, we had followed the particles until they reached the heliocentric distance $r=10~000$ au. Then, the particles were discarded from the simulation and their positions, ${\bf r}$, and velocities, ${\bf v}$, were recorded. For each recorded particle, the eccentricity vector
\begin{eqnarray}\label{ecc-vec}
{\bf e}~=~\left(\frac{\vert{\bf v}\vert^{2}}{GM}-\frac{1}{\vert{\bf r}\vert}\right){\bf r} - \frac{{\bf r}\cdot{\bf v}}{GM}{\bf v}
\end{eqnarray}
was calculated. $-{\bf e}$ vector was taken as the aphelion direction and the spherical Galactic coordinates of the aphelion were calculated.
The orbital integration was carried out in {\small REBOUND} \citep{RL12}. We used WHFast integrator \citep{RT15} with a timestep of $10^{4}$ years. The particles were followed for 2 Gyr.

\section{Simulation results and discussion}\label{sec:results4}

In our simulations in MD, we found characteristic imprint of the EFE on the distribution of aphelia of particles recorded at a given time (let us call these injected particles). 

\begin{figure}
\begin{center}
\resizebox{\hsize}{!}{\includegraphics{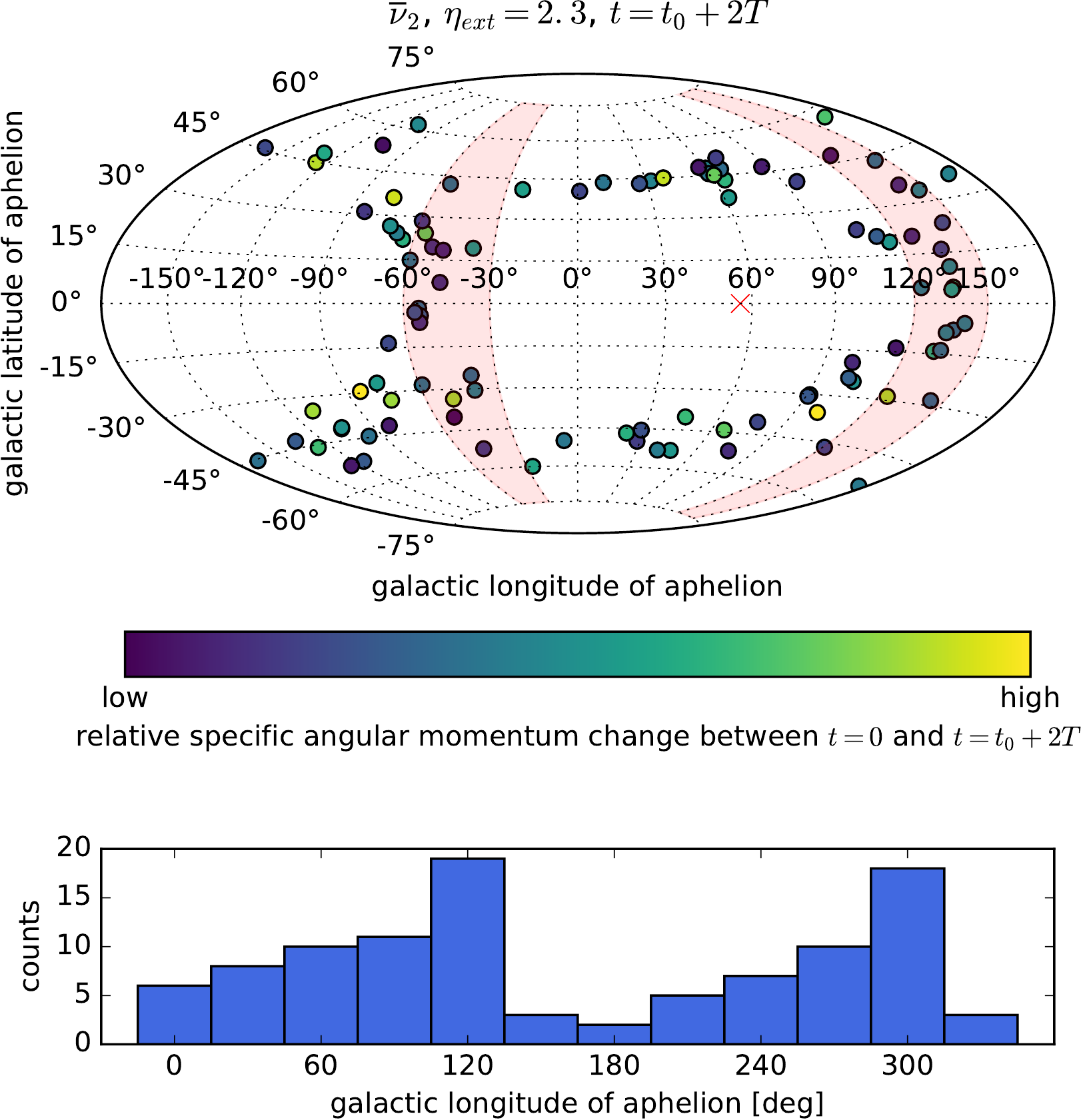}}
\caption{\textbf{Top.} Distribution of aphelia of simulated OC objects in the Galactic coordinates. The benchmark MD model incorporating the Galactic tides is considered. We show aphelia of all simulated OC bodies that came down to $r=10~000$ au during the time interval $t_{0}+2T\pm\Delta t$, where $t_{0}=190$ Myr, $T=230.8$ Myr, and $\Delta t=2$ Myr. During $2\Delta t$, the variation of the direction of the external field is small, hence the plot is basically a snapshot. The cross indicates the direction of the external field at time instant $t_{0}+2T$. The aphelia points are color coded according to the relative specific angular momentum change between $t = 0$ and $t = t_{0} + 2T$. The shaded regions indicate the great circle emerging in the observational data of the intermediate OOC. \textbf{Bottom.} Histogram distribution of the Galactic longitudes of aphelia. The distribution peaks around longitudes 135 and 315 (-45) deg at $t=t_{0}+CT$, where $C$ is an integer.}
\label{img:bench}
\end{center}
\end{figure}

\begin{figure*}
\begin{center}
\resizebox{\hsize}{!}{\includegraphics{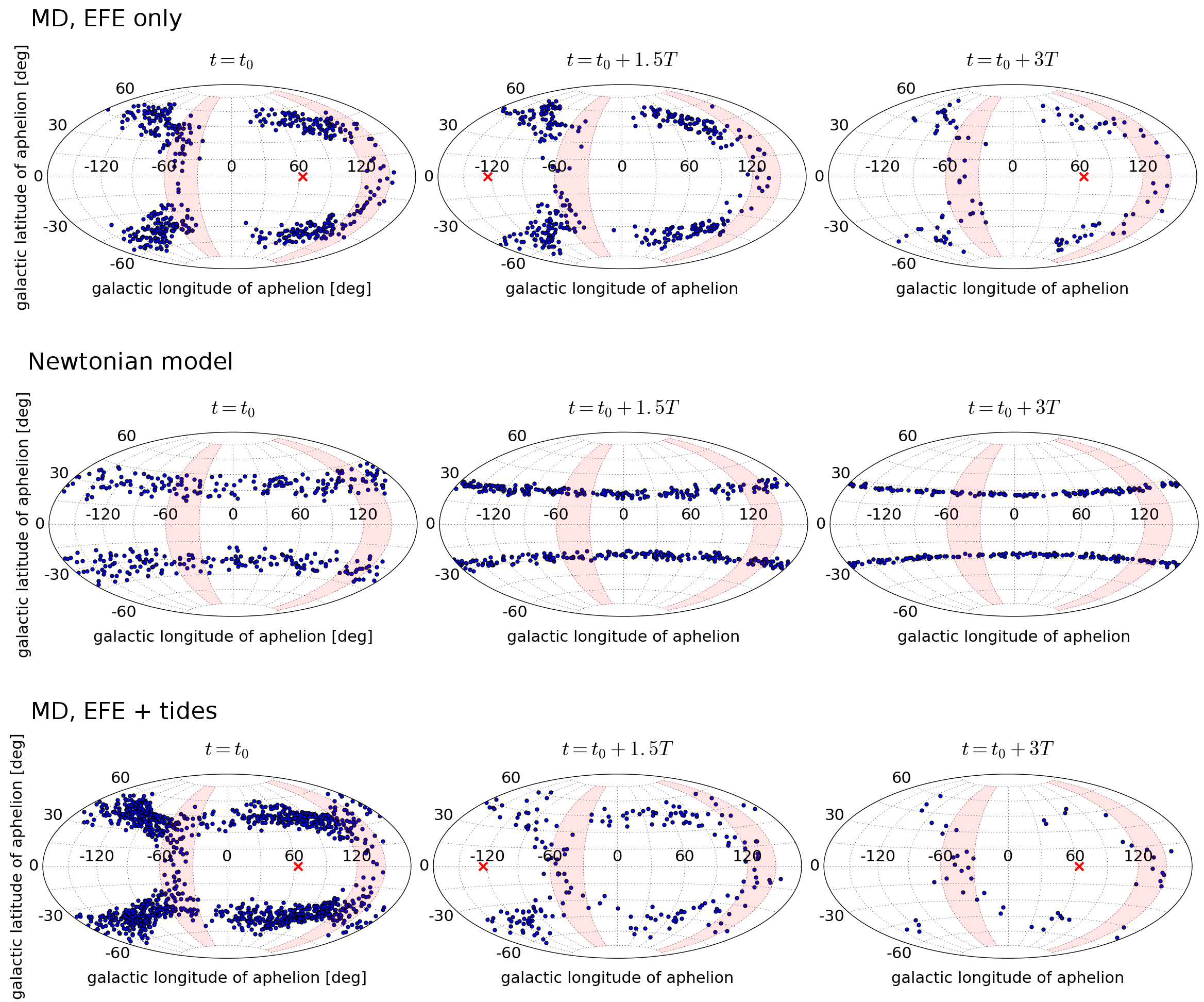}}
\caption{Distribution of aphelia of the simulated OC objects in the Galactic coordinates. The benchmark MD model without (first row) and with (third row) the Galactic tides is considered. The Newtonian, tides only, model is represented by the second row. The successive columns in each row show aphelia of all simulated OC bodies that came down to $r=10~000$ au during time intervals $t_{0}\pm\Delta t$, $t_{0}+1.5T\pm\Delta t$, and $t_{0}+3T\pm\Delta t$, where $t_{0}=190.0$ Myr was chosen to match the great circle concentration in the observational data, $T=230.8$ Myr is the period of the Sun's rotation around the Galactic center in the model, and $\Delta t=2$ Myr; the simulation starts at $t=0$ and ends at $t=2$ Gyr. Variation of the direction of the external field during $2\Delta t$ is negligible. The cross indicates direction of the external field at time instant $t_{0}$ (first column), $t_{0}+1.5T$ (second column), and $t_{0}+3T$ (third column). The external field direction (and the cross) moves from right to left as time goes. The shaded regions indicate the great circle emerging in the observational data.}
\label{img:astimegoes}
\end{center}
\end{figure*}

In Figs. \ref{img:bench} and \ref{img:astimegoes}, the distribution of aphelia in the Galactic coordinates is depicted for the injected particles.
When Newtonian, tides only, model is considered (middle row of Fig. \ref{img:astimegoes}), the aphelia cluster around $B=\pm30$ deg, without a preference in $L$, as expected \citep{R+08}. As time goes, the tidal infeed trajectories are continually less and less populous and the scatter of aphelia around the values $B=\pm30$ deg continually decreases. This is where the effect of passing stars would step in, repopulating the tidal infeed trajectories \citep{R+08}.

The MD benchmark model simulation shows an apparent non-uniformity not only in the distribution of $B$ but $L$ as well.
In Fig. \ref{img:astimegoes}, aphelia of the particles that came down to $r=10~000$ au during time intervals $t_{0}\pm\Delta t$, $t_{0}+1.5T\pm\Delta t$, and $t_{0}+3T\pm\Delta t$, where $T=230.8$ Myr is the period of the Sun's rotation around the Galactic center in the model, are shown. $\Delta t$ is set to $\Delta t=2$ Myr in order to obtain a meaningfully rich sample. We note that during $2\Delta t$, the variation of the external field direction is negligible. $t_{0}$ is the time corresponding to the phase at which the simulated great circle position corresponds to the observed one. 
The particles' aphelia cluster around a great circle in $L$. The clustering is the most apparent in the equatorial region.
The position of the great circle rotates with the rotating direction of the external field with period $T$. 
The trend is that after few $T$, the GCC in $L$ begins to stand out relatively to the horizontal stripes centered at $B=\pm45$ deg, and the simulated distribution resembles the observed one in Fig. \ref{img:data_tocomp}; see also Fig. \ref{img:bench}.
We tested that this picture does not depend on the heliocentric distance where we record the eccentricity vector (let us call it the injection distance); but we have to remain in the region where the gravity of the Sun is smaller than the external field, i.e. $r\gtrsim5500$ au. We repeated the simulations with the injection distance set also to 12 000 and 8000 au, getting qualitatively the same picture.

The simulated GCC matches the position of the observed one at the time when the external field is about 60 deg behind the present-day orientation. The external field needs additional about 38 Myr to align with the present-day orientation. As our model does not allow for tracing of the particles much below the heliocentric distance of 10 000 au, we are not able to determine the orientation of the external field at the time of an actual injection into the inner solar system. In any case, if the delineated position of the GCC is preserved, what is indicated by the fact that the dominant change in perihelion distance is made in the regions where our model applies, then this poses a problem for the EFE scenario.
Another complication is that we do not know on what timescale stellar encounters reshuffles the cloud in MD, if we presume that the effect of passing stars is similar as in Newtonian dynamics. Thus, we are not able to make definite quantitative prediction. 

The dependence on the interpolating function and the strength of the external field was tested. Restricting to combinations in Table \ref{tab:IF_tab}, there is no combination but the benchmark triplet $\overline{\nu}_{2}$, $a_{0}=0.815\times10^{-10}$ m s$^{-2}$, $\eta_{ext}=2.3$, that would yield a rotating GCC of aphelia of the injected particles. In all the other models, the equatorial region between about $B=30$ deg north and $B=30$ deg south is empty. In the case of $\widehat{\nu}_{6}$ and $\nu_{7}$, the aphelia distribution resembles that of the Newtonian, tides-only, model, with some concentrations and dilutions in the horizontal stripes, lying at $L$s whose positions change with time. In all the other remaining combinations, the effect of MD is too weak and is washed away by the Galactic tides. These findings are in concordance with what we find in \cite{Pau17} for Sednoids, i.e. we could elucidate both, the orbits of Sednoids and the GCC of intermediate-$a$ OOC comets, if we use parameters similar to those in the benchmark model. 

We also investigated the dependence on initial semimajor axis. If we initialise particles with lower energies, $X$ between 15 and 30 ($33000\lessdot~a\lessdot67000$ au), while leaving the procedure intact otherwise, the injected particles bear only weak imprint of MD. The aphelia distribution resembles that of the Newtonian, tides-only, model, with some barely-noticeable concentrations and dilutions in the horizontal stripes, lying at $L$s whose positions change with time. This is in line with the fact that the GCC, or any other form of $L$-non-uniformity, is not seen in the high-$a$ data.

\section{Summary}\label{sec:discussion}

Aphelia of intermediate OOC are distributed non-uniformly on the celestial sphere, showing an apparent concentration around the great circle centered at Galactic longitudes $L=-45$ and 135 deg. 
In the framework of MD, we have studied if there exist preferred orientations of orbits of OC bodies with intermediate semimajor axes where their perihelia are more effectively reduced. At heliocentric distances where the gravity of the Sun is much smaller than the external field of the Galaxy, i.e. for $r\gtrsim10~000$ au, MD potential incorporating EFE can be well approximated with the analytical formula of Eq. (\ref{QUMOND_EFE}).
Using Eq. (\ref{QUMOND_EFE}), we have shown analytically for orbits in the Galactic equatorial region that such preferred locations, though varying with time, indeed exist.
The fact that we do not observe non-uniformity also in the $L$ distribution of high-$a$ comets can be explained as the consequence of the fact that the influence of EFE weakens as $a$ increases, while the influence of the Galactic tides grows stronger rapidly with increasing $a$. We have backed up this claim by numerical experiments as well.

In numerical simulations, we demonstrated characteristic imprint of EFE on the distribution of aphelia of candidate OOC comets that migrated down to $r=10~000$ au.
EFE is able to induce GCC in $L$ for the intermediate-$a$ bodies. The appearance of the simulated GCC qualitatively agrees with the observed one.
The combined effect of EFE and tides leads to the distribution of aphelia resembling the observed one.
As the external field rotates around the Sun with period of one Galactic year, the longitude of the GCC of aphelia rotates with the same rate. 

We have found only one combination of Cassini-allowed interpolating function and Galactic parameters that leads to sufficiently strong EFE and as a consequence to the GCC in $L$; see the highlighted line in Table \ref{tab:IF_tab}. 
Though, note that the list of considered combinations is far from extensive. These findings are in concordance with the work of \cite{Pau17}, where we found that similarly strong EFE is needed to explain the large perihelion distances of Sedna and 2012 VP$_{113}$.

As a future prospect, it remains to be demonstrated that the distribution of aphelia of OC comets delivered into the inner solar system can be explained also quantitatively by the combined effect of EFE, Galactic tides, and stellar encounters in the framework of MD. An important question is also whether the GCC induced by EFE matches the position of the observed one at the time of the actual injection into the inner solar system when the external field has the present-day orientation.
To accomplish these tasks we would have to switch from the presented semi-analytical modeling to solving the QUMOND equations numerically.

\begin{acknowledgements}
I am thankful to Jozef Vil\'{a}gi and Leonard Korno\v{s} for providing a computational facility. 
Simulations in this paper made use of the \rebound code which can be downloaded freely at \url{http://github.com/hannorein/rebound}.
\end{acknowledgements}

%
%

\bibliographystyle{aa}
\bibliography{aniso} 

\begin{appendix}
\section{Secular variation of perihelion distance due to EFE}\label{sec:app4}

We aim to find the secular evolution of perihelion distance $q$ of an OC body, i.e. $\left\langle dq/dt \right\rangle=(1-e)\left\langle da/dt \right\rangle-a\left\langle de/dt \right\rangle$, driven by EFE in the region where the external fields dominates. Let $(R,T,N)$ be components of the perturbing acceleration due to the potential $\Phi$ in Eq. (\ref{QUMOND_EFE}), in a way that $R$ is a component in the direction of the heliocentric position vector ${\bf r}$, $T$ lies in the plane of the body's orbit and is perpendicular to $R$, and $N$ completes the right-handed coordinate system. For simplicity, we consider only bodies orbiting in the Galactic midplane.
Using the classical Gauss equations associated with the perturbed two-body problem Sun-OC body, we get the time variation of mutually tied-up orbital elements $a$, $e$, $\varpi$, where $\varpi$ is perihelion longitude, as:
\begin{eqnarray}\label{adddt}
\frac{da}{dt} & = & \frac{2a^{3/2}}{\mu^{1/2}\sqrt{1-e^{2}}}\left[R e\sin f + T(1+e\cos f)\right] \nonumber \\
\frac{de}{dt} & = & \frac{a^{1/2}\sqrt{1-e^{2}}}{\mu^{1/2}}\left[R \sin f + T\left(\cos f+\frac{e+\cos f}{1+e\cos f}\right)\right]\nonumber \\
\frac{d\varpi}{dt} & = & \frac{a^{1/2}\sqrt{1-e^{2}}}{\mu^{1/2}e}\left[-R \cos f + T\left(\frac{2+e\cos f}{1+e\cos f}\right)\sin f\right]
\end{eqnarray}
where
\begin{eqnarray}\label{ap1}
R & = & (-\nabla\Phi)\cdot {\bf e_{R}}\nonumber \\
T & = & (-\nabla\Phi)\cdot {\bf e_{T}}\nonumber \\
{\bf e_{R}} & = & [x /r,~y /r,~0] \nonumber \\
            & = & [\cos(\varpi+f),~\sin(\varpi+f),~0]\nonumber \\
{\bf e_{T}} & = & [-y /r,~x /r,~0] \nonumber \\
            & = & [-\sin(\varpi+f),~\cos(\varpi+f),~0]~.
\end{eqnarray}
For $\sin^{2}\theta$ in Eq. (\ref{QUMOND_EFE}), we can write on the basis of Eqs. (\ref{variest}) and (\ref{ap1}) that
\begin{eqnarray}\label{ap2}
\sin^{2}\theta & = & 1-\frac{\left[x \cos(\Omega_{0} t) + y \sin(\Omega_{0}t)\right]^{2}}{r^{2}}\nonumber \\
\Omega_{0} & < & 0~.
\end{eqnarray}
Finally, by using
\begin{eqnarray}\label{ap3}
\left\langle \frac{dc_{i}}{dt} \right\rangle & = & \frac{1}{P}\int_{0}^{P}\frac{dc_{i}}{dt}dt=\frac{1}{2\pi}\int^{2\pi}_{0}\frac{dc_{i}}{dt}
\frac{(1-e^{2})^{3/2}}{(1+e\cos f)^{2}}df\nonumber \\
\{c_{i}\}_{i=1,2,3} & = & \{a,e,\varpi\}\nonumber \\
L & = & \varpi+\pi
\end{eqnarray}
we arrive at
\begin{eqnarray}\label{dddt2}
\left\langle \frac{da}{dt} \right\rangle & = & 0 \nonumber \\
\left\langle \frac{de}{dt} \right\rangle & = & \mu^{1/2}K_{0}\nu_{ext}\frac{\sin [2(\vert\Omega_{0}\vert~t+L)]}{2a^{3/2}}F(e)~ \nonumber \\
\left\langle\frac{dL}{dt}\right\rangle & = & - \mu^{1/2}K_{0}\nu_{ext}\frac{\cos[2(\vert\Omega_{0}\vert~t+L)]}{2a^{3/2}}G(e)
\end{eqnarray}
where
\begin{eqnarray}\label{fege}
F(e) & = & \frac{2-2\sqrt{1-e^{2}}-e^{2}}{e^{4}}\frac{e^{3}-e}{\sqrt{1-e^{2}}} \nonumber \\
G(e) & = & \frac{2-2\sqrt{1-e^{2}}-e^{2}}{e^4}~.
\end{eqnarray}

\end{appendix}

\end{document}